# Causality and the Power Spectrum

James Robinson and Benjamin D. Wandelt

*Blackett Laboratory, Imperial College*

*Prince Consort Road, London SW7 2BZ, UK*

## Abstract

We find constraints on the generation of super-causal-horizon energy perturbations from a smooth initial state, under a simple physical scheme. We quantify these constraints by placing the upper limit $\lambda_{\rm c} \simeq 3.0 d_{\rm H}$ on the wavelength at which the power spectrum turns over to $k^4$ behavior. This means that sub-horizon processes can generate significant power on scales further outside the horizon than one might naively expect. The existence of this limit may have important implications for the interpretation of the small scale power spectrum of the Cosmic Microwave Background.

## I. INTRODUCTION

The two best candidate theories explaining the origin of structure in the universe are "defects" [1,2] and "inflation" [3]. There are many uncertainties in each theory, and the overlap between their predictions is considerable. In spite of this, the predictions of each theory are constrained, because any model must satisfy the laws of physics. In particular, any model is constrained by *causality*, which plays a fundamentally different role in either scenario.

This difference can be understood in terms of the evolution of two scales, the *Hubble distance* ($H^{-1}$) and the *causal horizon* ($d_{\rm H}$). The former is the crucial scale for the evolution of density perturbations (cf. Equation 82.9 in [4]). The latter is the largest scale on which causal interactions can seed density perturbations. We note that the word "horizon" has many uses in cosmology: The causal horizon we refer to here is the absolute limit for causal interactions. In other contexts, $H^{-1}$ can also be considered as a horizon, since it is the maximum distance over which causal interactions can occur during one expansion time.

In defect theories, $H^{-1}$ is of order $d_{\rm H}$ at all times. Making the usual assumption of a smooth initial state, super-horizon-distance and consequently super-Hubble-distance curvature perturbations must be suppressed. During a period of inflation on the other hand, $d_{\rm H}$ grows to some large value, while $H^{-1}$ remains constant. This means that $d_{\rm H} \gg H^{-1}$ for the subsequent history of the universe, and therefore that super-Hubble curvature perturbations can exist. (Cf. Figure 8.4 in [5]).

Since the Hubble distance is the crucial scale for the evolution of density perturbations, the presence or absence of super-Hubble curvature perturbations may have profound observable consequences. If so, it will be possible to distinguish between defects and inflation in spite of the lack of preferred models. One way to do this has been proposed by Albrecht et al. [6]: It is well known that inflation gives rise to oscillations in the small scale power spectrum of the Cosmic Microwave Background [3]. They predict that these oscillations will not be present for defect theories, provided energy fluctuations are sufficiently suppressed on super-Hubble scales[1]. It is the issue of quantifying this suppression which we address in this paper.

For generality, we phrase our discussion in terms of the horizon distance $d_{\rm H}$. It should therefore apply to any theory satisfying the laws of physics which starts from a smooth initial state. The key quantity is the "energy" $\tau_{00}$, where $\tau_{\mu\nu}$ is the stress energy pseudotensor used in [8] and [9]. $\tau_{\mu\nu}$ is particularly useful, because it describes the flow of energy and momentum in the universe in an intuitive way (cf. Section II) and because $\tau_{00}$ is the coefficient of the growing mode of matter perturbations. The power spectrum $P(\boldsymbol{k})$ is defined via

$$\langle \tilde{\tau}_{00}(\boldsymbol{k}) \tilde{\tau}^*_{00}(\boldsymbol{k}') \rangle = (2\pi)^3 \; P(\boldsymbol{k}) \; \delta^{(3)}(\boldsymbol{k} - \boldsymbol{k}') \tag{1}$$

where $\delta^{(3)}$ denotes the Dirac delta in three dimensions. We use the convention

$$\tilde{f}(\boldsymbol{k}) = \frac{1}{(2\pi)^{3/2}} \int d^3\boldsymbol{x} \; e^{-i\boldsymbol{k}.\boldsymbol{x}} \; f(\boldsymbol{x}) \tag{2}$$

The nature of the constraint on $P(k)$ for scales far outside the causal horizon is well known. Several authors [9–13] showed that $P(k)$ must fall as $k^4$ in this limit. However, knowing this large scale limiting behavior is not enough. In addition, we need to know where the $k^4$ behavior sets in. We can quantify this in terms of a turnover wave-number $k_{\rm c} = \frac{2\pi}{\lambda_{\rm c}}$. For any power spectrum with white noise behavior on small scales and $k^4$ behavior on large scales we can define this turnover by writing the limits as follows[2].

$$P(k) = \begin{cases} C & \ldots k \gg {d_{\rm H}}^{-1} \\ C\left(\frac{k}{k_c}\right)^4 & \ldots k \ll {d_{\rm H}}^{-1} \end{cases} \tag{3}$$

---

[1]This is an active area of research. Numerical simulations described in [7] suggest that in some defect models the suppression of curvature perturbations may alter rather than annihilate these oscillations.

[2]$k_{\rm c}$ defined here corresponds directly to that used by Albrecht and Stebbins in [14], since their "compensation" factor $\left(\frac{1}{1+(k/k_{\rm c})^2}\right)^2$ has exactly the same limits.

More general perturbations need not have white noise behavior on small scales, but $k_c$ can still be defined in a similar way.

Given a specific defect model, we should be able to work out a value for $k_c$. In the absence of specific models, we may still be able to place a lower limit on it. In this paper, we work out a lower limit given one simple scenario for causal perturbations, where matter undergoes random displacements at each point in space. Although we do not consider a general case, we expect that all other methods of perturbing matter will produce power spectra with the same or higher values of $k_c$.

The paper is set out as follows. In Section II we set up a formalism for generating perturbations in a fluid, subject to the laws of causality and energy and momentum continuity. In Section III we use this formalism to derive a minimum for $k_c$ in our simple scenario, and consider the relevance of this limit to more general perturbations. In Section IV we draw our conclusions.

## II. GENERAL FORMALISM AND $K^4$ BEHAVIOR

We begin with a prescription for generating "causal" fluctuations in the energy. We note that the stress-energy pseudotensor satisfies an ordinary (non-covariant) conservation law. That is

$$\partial_\nu \tau_{\mu\nu} = 0 \tag{4}$$

The zeroth component is the energy continuity equation. This implies that changes in energy from a smooth initial state can only arise as a result of displacing energy elements. Initially, consider a homogeneous field with energy $\tau_{00}(\boldsymbol{x}) = 0 \;\forall\; \boldsymbol{x}$. We then move $n$ energy elements with weight $m_i$ from $\boldsymbol{x}$ to the new positions $\boldsymbol{x} + \boldsymbol{\Delta}_i(\boldsymbol{x})$. Doing this at every point in space gives the new energy

$$\tau_{00}(\boldsymbol{x}) = \int d^3\boldsymbol{x}' \Big[\sum_{i=1}^{n} m_i \delta^{(3)}\Big(\boldsymbol{x} - \boldsymbol{x}' - \boldsymbol{\Delta}_i(\boldsymbol{x})\Big) - M\delta^{(3)}\Big(\boldsymbol{x} - \boldsymbol{x}'\Big)\Big] \tag{5}$$

where $M = \sum m_i$. We imagine perturbations occurring in this way as it is then possible to ensure that they also satisfy other physical laws. At the very least, physics requires

1. *Momentum Continuity:* We can generate any $\tau_{00}(\boldsymbol{x})$ arising via a process satisfying momentum continuity by choosing a set of displacement fields $\boldsymbol{\Delta}_i(\boldsymbol{x})$ satisfying

$$\sum_{i=1}^{N} m_i \boldsymbol{\Delta}_i(\boldsymbol{x}) = 0 \tag{6}$$

2. *Causality:* Causality implies constraints on energy perturbations which can be most simply stated by saying that no energy element can move further than the causal horizon. That is,

$$|\boldsymbol{\Delta}_i(\boldsymbol{x})| \leq d_{\rm H} \tag{7}$$

Substituting from Equation 5 into Equation 2 we obtain

$$\tilde{\tau_{00}}(\boldsymbol{k}) = \frac{1}{(2\pi)^{3/2}} \int d^3\boldsymbol{x}\; e^{-i\boldsymbol{k}.\boldsymbol{x}} \sum_{i=1}^{n} m_i \left( e^{-i\boldsymbol{k}.\boldsymbol{\Delta}_i(\boldsymbol{x})} - 1 \right) \tag{8}$$

For completeness, we demonstrate the well known result [9–13] that $k^4$ behavior follows from these assumptions. For modes well outside the horizon we have $\boldsymbol{k}.\boldsymbol{\Delta}_i \ll 1$. Taylor expanding the second exponential we find

$$\tilde{\tau_{00}}(\boldsymbol{k}) = \frac{1}{(2\pi)^{3/2}} \int d^3\boldsymbol{x}\; e^{-i\boldsymbol{k}.\boldsymbol{x}} \left( -i \sum_{i=1}^{n} m_i \boldsymbol{k}.\boldsymbol{\Delta}_i(\boldsymbol{x}) - \frac{1}{2} \sum_{i=1}^{n} m_i \left(\boldsymbol{k}.\boldsymbol{\Delta}_i(\boldsymbol{x})\right)^2 + \mathcal{O}(k^3) \right) \tag{9}$$

Momentum conservation (Equation 6) ensures that first term in this expansion vanishes. The leading term is therefore the one proportional to $k^2$, implying that the power spectrum must go like $k^4$.

Next, we use the formalism developed here to work out a minimum for the *turnover* in the power spectrum under a simple scheme for perturbing matter.

## III. TURNOVER OF POWER SPECTRUM

In this section, we consider the simplest scheme for perturbing matter which satisfies the laws of physics as stated in Section II. In this scheme, two energy elements with the same weight are displaced equally but in opposite directions from each point in space. Further, the displacement vectors of energy elements at each point in space are independent. The index $i$ then runs from 1 to 2, with $\boldsymbol{\Delta}_1 = -\boldsymbol{\Delta}_2 = \boldsymbol{\Delta}$. Equation 8 reduces to

$$\tilde{\tau_{00}}(\boldsymbol{k}) = \frac{A}{(2\pi)^{3/2}} \int d^3\boldsymbol{x}\ e^{-i\boldsymbol{k}.\boldsymbol{x}}\ \left[\cos \boldsymbol{k}.\boldsymbol{\Delta}(\boldsymbol{x}) - 1\right] \tag{10}$$

where $A$ is some constant. We want to work out the power spectrum, defined in Equation 1. We have

$$\tilde{\tau_{00}}(\boldsymbol{k})\tilde{\tau_{00}}^*(\boldsymbol{k}') = \frac{A^2}{(2\pi)^3} \int d^3\boldsymbol{x}\ d^3\boldsymbol{x}'\ e^{-i(\boldsymbol{k}.\boldsymbol{x}-\boldsymbol{k}'.\boldsymbol{x}')}\ \left[\cos \boldsymbol{k}.\boldsymbol{\Delta}(\boldsymbol{x}) - 1\right]\left[\cos \boldsymbol{k}'.\boldsymbol{\Delta}(\boldsymbol{x}') - 1\right] \tag{11}$$

To obtain the expectation value of this quantity, we must perform the functional integration

$$\langle \tilde{\tau_{00}}(\boldsymbol{k})\tilde{\tau_{00}}^*(\boldsymbol{k}')\rangle = \int [d\boldsymbol{\Delta}]\ \ P\left[\boldsymbol{\Delta}\right]\ \ \tilde{\tau_{00}}(\boldsymbol{k})\ \tilde{\tau_{00}}^*(\boldsymbol{k}') \tag{12}$$

where $P[\boldsymbol{\Delta}]$ is the probability functional of obtaining a displacement field $\boldsymbol{\Delta}$. In the case where the displacement at each point in space is independent, we can separate $P[\boldsymbol{\Delta}]$ into a product of probability distributions for the displacement at each point in space. That is

$$P\left[\boldsymbol{\Delta}\right] = \prod_i\ p\left(\boldsymbol{\Delta}_i\right) \tag{13}$$

where we have discretized space so that the index $i$ takes $\boldsymbol{x}_i$ through all points and $\boldsymbol{\Delta}_i$ is the value of the field at the point $\boldsymbol{x}_i$. Using this discretization, and taking $dV$ to be the volume of each spatial element, we get

$$\langle \tilde{\tau_{00}}(\boldsymbol{k})\tilde{\tau_{00}}^*(\boldsymbol{k}')\rangle = \\ \frac{A^2}{(2\pi)^3} \int \prod_i \left[d^3\boldsymbol{\Delta}_i p\left(\boldsymbol{\Delta}_i\right)\right] \sum_{j,k} (dV)^2 e^{-i(\boldsymbol{k}.\boldsymbol{x}_j-\boldsymbol{k}'.\boldsymbol{x}_k)}\ \left[\cos \boldsymbol{k}.\boldsymbol{\Delta}_j - 1\right]\left[\cos \boldsymbol{k}'.\boldsymbol{\Delta}_k - 1\right] \tag{14}$$

For the $j,k$th element in the sum, all but the $j$th and $k$th integrals can be carried out trivially, giving the value 1. Hence we obtain

$$\langle \tilde{\tau_{00}}(\boldsymbol{k})\tilde{\tau_{00}}^*(\boldsymbol{k}')\rangle = \frac{A^2}{(2\pi)^3} \sum_{j,k} (dV)^2 \\ \times \int d^3\boldsymbol{\Delta}_j p\left(\boldsymbol{\Delta}_j\right) \int d^3\boldsymbol{\Delta}_k p\left(\boldsymbol{\Delta}_k\right) e^{-i(\boldsymbol{k}.\boldsymbol{x}_j-\boldsymbol{k}'.\boldsymbol{x}_k)}\ \left[\cos \boldsymbol{k}.\boldsymbol{\Delta}_j - 1\right]\left[\cos \boldsymbol{k}'.\boldsymbol{\Delta}_k - 1\right] \tag{15}$$

Separating the sum into parts with $j \neq k$ and $j = k$, and taking the continuum limit, we get

$$
\begin{aligned}
\langle \tilde{\tau_{00}}(\boldsymbol{k})\tilde{\tau_{00}}^*(\boldsymbol{k}')\rangle = & \frac{A^2}{(2\pi)^3}\left|\int d^3\boldsymbol{x}\int d^3\boldsymbol{\Delta}\; p\,(\boldsymbol{\Delta})\, e^{-i\boldsymbol{k}.\boldsymbol{x}}\;\left[\cos\boldsymbol{k}.\boldsymbol{\Delta}-1\right]\right|^2 \\
& +\frac{A^2}{(2\pi)^3}\int d^3\boldsymbol{x}d^3\boldsymbol{x}'\delta^{(3)}(\boldsymbol{x}-\boldsymbol{x}')e^{-i(\boldsymbol{k}.\boldsymbol{x}-\boldsymbol{k}'.\boldsymbol{x}')} \\
& \times\int d^3\boldsymbol{\Delta}d^3\boldsymbol{\Delta}'\delta^{(3)}(\boldsymbol{\Delta}-\boldsymbol{\Delta}')p\,(\boldsymbol{\Delta})\,p\,(\boldsymbol{\Delta}')\left[\cos\boldsymbol{k}.\boldsymbol{\Delta}-1\right]\left[\cos\boldsymbol{k}'.\boldsymbol{\Delta}'-1\right]
\end{aligned}
\tag{16}
$$

The first term vanishes, leaving

$$
\langle \tilde{\tau_{00}}(\boldsymbol{k})\tilde{\tau_{00}}^*(\boldsymbol{k}')\rangle = A^2\,\delta^{(3)}(\boldsymbol{k}-\boldsymbol{k}')\int d^3\boldsymbol{\Delta}\; p^2\,(\boldsymbol{\Delta})\;\left[\cos\boldsymbol{k}.\boldsymbol{\Delta}-1\right]^2 \tag{17}
$$

The power spectrum is then just

$$
P(\boldsymbol{k}) = \frac{A^2}{(2\pi)^3}\int d^3\boldsymbol{\Delta}\; p^2\,(\boldsymbol{\Delta})\;\left[\cos\boldsymbol{k}.\boldsymbol{\Delta}-1\right]^2 \tag{18}
$$

We can simplify this expression by assuming that the probability distribution is isotropic, which implies that $P(\boldsymbol{k}) = P(|k|)$. Choosing $\boldsymbol{k} = (0, 0, k)$ we find

$$
P(k) = \frac{A^2}{(2\pi)^3}\int dz\;\mathcal{P}(z)\;\left[1-\cos kz\right]^2 \tag{19}
$$

where

$$
\begin{aligned}
\mathcal{P}(\Delta_z) &= \int d\Delta_x\; d\Delta_y\; p^2(\boldsymbol{\Delta}) \\
&= \int_{|\Delta_z|}^{\infty} d\Delta_r\;\int_0^{2\pi} d\Delta_\phi\; p^2(\boldsymbol{\Delta})
\end{aligned}
\tag{20}
$$

and $\Delta_z$,$\Delta_r$ and $\Delta_\phi$ are the components of $\boldsymbol{\Delta}$ in cylindrical polar coordinates. If no energy element can move further than some horizon distance $d_{\mathrm{H}}$, then $p(\boldsymbol{\Delta})$ must vanish for $|\boldsymbol{\Delta}| > d_{\mathrm{H}}$, and $\mathcal{P}(z)$ must vanish for $z > d_{\mathrm{H}}$. We then see that the limiting behavior for $P(k)$ is

$$
P(k) = \begin{cases} \frac{A^2}{(2\pi)^3}\;\int dz\;\mathcal{P}(z) & \ldots k \gg {d_{\mathrm{H}}}^{-1} \\ \frac{A^2}{(2\pi)^3}\;\frac{k^4}{4}\int dz\; z^4\;\mathcal{P}(z) & \ldots k \ll {d_{\mathrm{H}}}^{-1} \end{cases} \tag{21}
$$

Comparing the limiting behavior with the form described in Equation 3, we see that the turnover of the power spectrum occurs at

$$k_c = \left( \frac{\int dz\ \mathcal{P}(z)}{\frac{1}{4}\int dz\ z^4\ \mathcal{P}(z)} \right)^{\frac{1}{4}} \tag{22}$$

Since the integrand in Equation 20 is positive definite, we see that $\mathcal{P}(z)$ must be a constant or monotonically decreasing function of $z$. By inspection of Equation 22 we see that $k_c$ is minimized if we take $\mathcal{P}(z)$ to be a constant for $0 \leq z \leq d_{\mathrm{H}}$. This corresponds to the following choice for the probability distribution $p(\mathbf{\Delta})$:

$$p^2(\mathbf{\Delta}) \propto \delta^{(1)}(\Delta_r - d_{\mathrm{H}}) \tag{23}$$

This is the extreme case where the energy elements always move as far as they are allowed. Substituting $\mathcal{P}(z)$ into Equation 19 and carrying out the integration, we find the explicit solution

$$P(k) \propto \left( 1 - \frac{4 \sin k d_{\mathrm{H}}}{3 k d_{\mathrm{H}}} + \frac{\sin 2 k d_{\mathrm{H}}}{6 k d_{\mathrm{H}}} \right) \tag{24}$$

By examining the limiting behavior, we find that the value of $k_c$ is $20^{\frac{1}{4}} {d_{\mathrm{H}}}^{-1} \simeq 2.1 {d_{\mathrm{H}}}^{-1}$. The corresponding wavelength is $\lambda_c \simeq 3.0 d_{\mathrm{H}}$.

In Figure 1 we compare this power spectrum to the results of a computer simulation where we perturb matter on a $128 \times 128 \times 128$ lattice by moving pairs of energy elements randomly from each point through a distance $d_{\mathrm{H}}$. We note that there is a very good agreement between the theoretical and the simulated power spectra, up to deviations for large $k$ which are finite grid effects. This agreement represents a solid check on our calculation. Further, it is interesting to compare the continuum calculation described in this paper with the discrete case represented by the computer simulation. The discrete case does not involve the unphysical assumption that matter at infinitesimally close points in space undergoes displacements in arbitrarily different directions. Figure 1 illustrates that this only affects the small scale region of the power spectrum, and not its turnover.

In this section, we have worked out limits on the generation of energy perturbations, given minimal assumptions about causality and energy and momentum continuity. Although we have only considered special perturbations, we believe that the value of $k_c$ we have found is a general minimum. To show why, we consider the two ways in which we could make our perturbations more general.

- Firstly, we could allow more than two energy elements, or energy elements of different weight, to be displaced from each point in space. Computer simulations suggest that this leaves $k_c$ unchanged or slightly increased.

- Secondly, we could introduce correlations between the displacements of energy elements at different points in space. Without performing any detailed calculations, we make the following observation. To introduce correlations, we must transport information between points which are separated in space. This transport of information takes time, which must be subtracted from the time available for energy elements to move. We expect that the net effect will be to increase $k_c$.

## IV. CONCLUSIONS

We have found constraints on super-horizon energy perturbations which can be generated from a smooth initial state, under a simple physical scheme. We have quantified these constraints in terms of an extremal power spectrum (Equation 24, Figure 1). Its turnover to $k^4$ behavior occurs at a wavelength $\lambda_c \simeq 3.0 d_{\rm H}$ or equivalently, at a wavenumber $k_c \simeq 2.1 d_{\rm H}^{-1}$. Hence we find that sub-Horizon processes can generate significant power on scales further outside the horizon than one might naively expect. Although we have only considered special perturbations, we have stated reasons for believing that this limit is general.

For any theory, constraints on super-Hubble perturbations can be inferred from constraints on super-horizon perturbations, provided we know how $H^{-1}$ and $d_{\rm H}$ evolve. Essentially, super-Hubble energy perturbations are suppressed for defect theories, and not for inflation, but the details will be model-dependent. Since the Hubble distance is the crucial scale for the evolution of density perturbations, any difference in this super-Hubble behavior may have profound observable consequences, for instance in the small scale power spectrum of the Cosmic microwave background [6]. The limit we have found in this paper makes it possible to quantify this without referring to specific models. Therefore, it may be possible to discover whether structure in the universe was seeded by defects or inflation, in spite of considerable uncertainties in both theories.

## ACKNOWLEDGMENTS

We are grateful to A. Albrecht and P. Ferreira for helpful discussions. J. R. is supported by P.P.A.R.C. Award No. 94313015. B. D. W. is supported by the Blackett Laboratory and H. & C. Wandelt.

# FIGURES

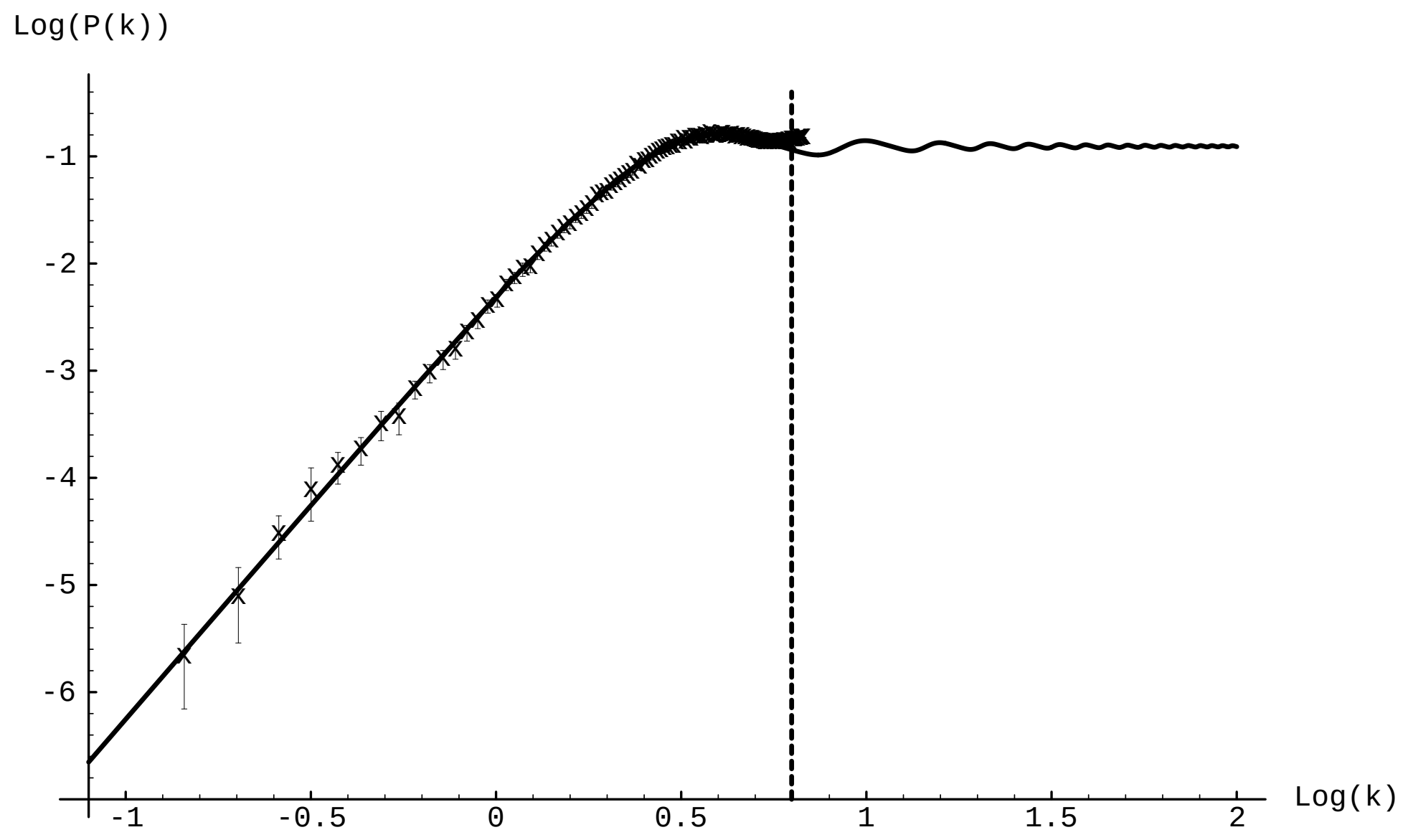

FIG. 1. Theoretical (solid line) and simulated (data points) extremal power spectrum. Power is measured in arbitrary units and $k$ in units of $d_{\mathrm{H}}^{-1}$. The vertical line corresponds to a mode which exactly fills the horizon ($kd_{\mathrm{H}} = 2\pi$).